\documentclass[3p,times,twocolumn]{elsarticle}
 \biboptions{comma,sort&compress}
 
 \usepackage{amsmath}
\usepackage{graphicx}
\usepackage{here}
\usepackage{ecrc}

\newcommand{\bea}{\begin{eqnarray}} \newcommand{\eea}{\end{eqnarray}}

\def\Comment#1{}

\newcommand{\bean}{\begin{eqnarray*}}
\newcommand{\eean}{\end{eqnarray*}}

\newcommand{\gapproxeq}{\lower
.7ex\hbox{$\;\stackrel{\textstyle >}{\sim}\;$}}
\newcommand{\lapproxeq}{\lower
.7ex\hbox{$\;\stackrel{\textstyle <}{\sim}\;$}}

\newcommand\lsim{\mathrel{\rlap{\lower4pt\hbox{\hskip1pt$\sim$}}
    \raise1pt\hbox{$<$}}}
\newcommand\gsim{\mathrel{\rlap{\lower4pt\hbox{\hskip1pt$\sim$}}
    \raise1pt\hbox{$>$}}}
\newcommand{\ba}{\begin{array}}
\newcommand{\ea}{\end{array}}
\newcommand{\nn}{\nonumber}

\newcommand{\be}{\begin{equation}}
\newcommand{\ee}{\end{equation}}
\newcommand{\bear}{\begin{eqnarray}}
\newcommand{\eear}{\end{eqnarray}}

\newcommand{\ket}{\,\rangle}
\newcommand{\bra}{\langle \,}

\newcommand{\Frac}[2]{\frac{\displaystyle #1}{\displaystyle #2}}
\newcommand{\Int}{\displaystyle{\int}}

\volume{00}

\firstpage{1}

\journalname{Nuclear and Particle Physics Proceedings}

\runauth{Ignasi Rosell}

\jid{nppp}

\jnltitlelogo{Nuclear and Particle Physics Proceedings}

\usepackage{amssymb}

\usepackage[figuresright]{rotating}

\begin{document}

\begin{frontmatter}

\title{Heavy resonances and the oblique parameters $S$ and $T$
 $^*$}
 \cortext[cor0]{Talk given at the 26th International Conference in Quantum Chromodynamics (QCD 23),  10-14 July 2023, Montpellier (France).}
 \author{Ignasi Rosell\corref{cor2}}
\ead{rosell@uchceu.es}
\address{Departamento de Matem\'aticas, F\'\i sica y Ciencias Tecnol\' ogicas, Universidad Cardenal Herrera-CEU, CEU Universities, 46115 Alfara del Patriarca, Val\`encia, Spain}
 \author{Antonio Pich}
\ead{pich@ific.uv.es}
\address{IFIC, Universitat de Val\`encia -- CSIC, Apt. Correus 22085, 46071 Val\`encia, Spain}
 \author{Juan Jos\'e Sanz-Cillero}
    \ead{jjsanzcillero@ucm.es}

\address{Departamento de F\'\i sica Te\'orica and IPARCOS,  Universidad Complutense de Madrid, E-28040 Madrid, Spain}
 \cortext[cor2]{Speaker, corresponding author.}

\pagestyle{myheadings}
\markright{ }
\begin{abstract}
It has been confirmed experimentally the existence of a mass gap between Standard Model (SM) and eventual Beyond Standard Model (BSM) fields. Therefore, the use of  effective approaches to search for fingerprints of New Physics is very appealing. A non-linear realizations of the Electroweak Symmetry Breaking is considered here, where the Higgs is a singlet with free couplings and the SM fields are also coupled to bosonic heavy resonances. A one-loop-level calculation of the oblique $S$ and $T$ parameters is presented here. This analysis allows us to constrain resonance masses to be above the TeV scale, $M_R\!\gsim\! 3\,$TeV, in good agreement with our previous determinations, where these observables were computed with a more simplified Lagrangian.\end{abstract}

\end{frontmatter}

\section{Introduction}

The LHC results have confirmed the Standard Model (SM) as the correct framework for explaining electroweak interactions within the energy ranges examined up to now. The discovery of a Higgs-like particle~\cite{higgs}, with couplings very aligned with SM predictions, has effectively filled in the complete set of fundamental fields according to the SM, and there have been no new states discovered to date. Consequently, the available data indicate the presence of a mass gap between the SM and any hypothetical degrees of freedom associated with New Physics (NP). This gap serves as a rationale for employing effective field theories to systematically investigate low-energy physics for indications of NP scales.

Effective field theories are built upon several key components, being the particle content, the symmetries and the power counting the most important ones. In the context of electroweak theory, the choice of power counting method hinges on how the Higgs field  $h$ is introduced~\cite{Buchalla:2016bse,Pich:2018ltt}. There are two different approaches: the commonly used linear realization of Electroweak Symmetry Breaking (EWSB), where the Higgs is considered part of a doublet together with  the three electroweak (EW) Goldstone bosons $\vec{\varphi}$, as in the SM; or the more encompassing non-linear realization, which does not presume any specific relationship between the Higgs and the three Goldstone fields. Here we opt for the latter approach~\cite{lagrangian}, employing an expansion based on generalized momenta. It is worth noting that the linear realization can be seen as a special case within the broader framework of the non-linear approach.

We examine a strongly-coupled scenario involving heavy spin-1 resonances interacting with the SM particles. Our objective is to evaluate the bounds placed on the masses of these heavy resonances based on the phenomenology derived from the electroweak oblique parameters~\cite{Peskin_Takeuchi}. 

In Ref.~\cite{ST} we presented a one-loop calculation of the $S$ and $T$ parameters within strongly-coupled scenarios of this kind. Our analysis yielded robust and model-independent constraints on both the Higgs couplings and the heavy scales involved. By making only modest assumptions about the high-energy behavior of the underlying fundamental theory, we demonstrated that precision electroweak data necessitate the Higgs-like scalar to possess a $hWW$ coupling very close to the SM one. Simultaneously, we found that the masses of vector and axial-vector resonances should exhibit considerable degeneracy and be above $4\,$TeV.  The considerably larger dataset collected in recent years has provided a much more precise experimental measurement of the $hWW$ coupling, $\kappa_W$. Consequently, we no longer treat $\kappa_W$ as an independent free parameter, opting instead to use its experimentally determined value as an input. This adjustment enables us to broaden the scope of our analysis to include a wider range of interactions: while our prior work solely focused on the bosonic P-even sector~\cite{ST}, we incorporate now both P-even and P-odd operators into our approach. Here we show the initial phase in the effort to enhance and update the results reported in Ref.~\cite{ST}; we plan to complete this update by incorporating fermionic contributions as well~\cite{future}.

In Section 2, we introduce our Lagrangian including the resonances, while Section 3 outlines the calculation of the $S$ and $T$ parameters, up to next-to-leading order (NLO), through dispersive representations. Section 4 is dedicated to exploring the practical implications of these results from a phenomenological perspective. Ultimately, we provide a summary in Section 5.

\section{The theoretical framework}

\subsection{The effective resonance Lagrangian}

In the non-linear realization of EWSB, operators are not ordered following their canonical dimensions and they must be organized taking into account their behavior at low momenta, the chiral dimensions~\cite{Weinberg}. Here we focus on the NLO resonance contributions to the $S$ and $T$ parameters derived exclusively from the lightest bosonic absorptive cuts ($\varphi\varphi$ and $h\varphi$).
Consequently, we only need to consider bosonic operators involving at most one spin-1 resonance field. Following the notation of Ref.~\cite{lagrangian}, the relevant CP-conserving Lagrangian for our calculations reads
\begin{eqnarray}
\Delta \mathcal{L}_{\mathrm{RT}} =
 \frac{v^2}{4}\,\left( 1 \!+\!\Frac{2\kappa_W}{v} h \right) \bra u_\mu u^\mu\ket_{2} \qquad \qquad \quad \quad  \nonumber \\ 
 +\,\bra V^1_{3\,\mu\nu} \left( \Frac{F_V}{2\sqrt{2}}  f_+^{\mu\nu} \!+\! \Frac{i G_V}{2\sqrt{2}} [u^\mu, u^\nu]  \!+\! \Frac{\widetilde{F}_V }{2\sqrt{2}} f_-^{\mu\nu}    \right.  \nonumber \\    \left.+ \Frac{ \widetilde{\lambda}_1^{hV} }{\sqrt{2}}\left[  (\partial^\mu h) u^\nu\!-\!(\partial^\nu h) u^\mu \right]    \right) \ket_2  \nonumber \\
  + \,\bra A^1_{3\,\mu\nu} \left(\Frac{F_A}{2\sqrt{2}}  f_-^{\mu\nu}  \!+\! \Frac{ \lambda_1^{hA} }{\sqrt{2}} \left[ (\partial^\mu h) u^\nu\!-\!(\partial^\nu h) u^\mu \right]  \right.  \nonumber \\  \left. +  \Frac{\widetilde{F}_A}{2\sqrt{2}} f_+^{\mu\nu} \!+\!  \Frac{i \widetilde{G}_A}{2\sqrt{2}} [u^{\mu}, u^{\nu}]    \right) \ket_2   
\,,  \label{Lagrangian}
\end{eqnarray} 
\label{eq:Lagr}
where $V^1_{3\,\mu\nu}$ and $A^1_{3\,\mu\nu}$ introduce color-singlet custodial-triplet resonances with $J^{PC}$ quantum numbers $1^{--}$ (V) and $1^{++}$ (A) by using an antisymmetric formalism. The Goldstone fields are parametrized through the SU(2) matrix $U = u^2 = \exp{(i \vec{\sigma}\vec{\varphi}}/v)$, where $v= (\sqrt{2} G_F)^{-1/2}=246\;\mathrm{GeV}$ is the EWSB scale,  $u_\mu = -i u^\dagger D_\mu U u^\dagger$ with $D_\mu$ the appropriate covariant derivative, and $f_\pm^{\mu\nu}$ contain the gauge-boson field strengths. Note that couplings with tilde are related to odd-parity operators.

\subsection{Short-distance constraints}

From (\ref{Lagrangian}) one observes the presence of ten resonance parameters (including masses), so that the use of asymptotic constraints is fundamental to reduce the number of unknown parameters and, consequently, to be able to obtain phenomenological bounds. Moreover, this resonance Lagrangian is assumed to be an interpolation between the low- and the high-energy regimes and, as a result, considering a reasonable high-energy behavior is an important ingredient of this effective approach:
\begin{enumerate}
\item Requiring that two-Goldstone ($\varphi \varphi$) and Higgs-Goldstone ($h \varphi$) vector and axial form factors vanish at high energies allow us to determine the couplings $G_V$, $\widetilde G_A$, $\lambda_1^{hA}$ and $\widetilde{\lambda}_1^{hV}$ in terms of the remaining parameters~\cite{PRD2},
\begin{equation}
\!\!\!\!\!\!\!\!\!\!\!\frac{G_V}{F_A} \!=\! - \frac{\widetilde G_A}{\widetilde{F}_V} \!=\! \frac{\lambda_1^{hA} v }{\kappa_W F_V} \!=\! -\frac{\widetilde{\lambda}_1^{hV} v}{\kappa_W \widetilde{F}_A} \!=\! \frac{v^2}{F_V F_A \!-\! \widetilde{F}_V \widetilde{F}_A} . \label{constraints_summary}
\end{equation}
\item The assumed chiral symmetry of the underlying electroweak theory suggests that the $W^3B$ correlator is an order parameter of the EWSB and it vanishes at high energies  in asymptotically-free gauge theories as $1/s^3$~\cite{Bernard:1975cd}, giving rise to the 1st and 2nd Weinberg Sum Rules (WSRs)~\cite{WSR}:
\begin{enumerate}
\item 1st WSR (vanishing of the $1/s$ term). At leading-order (LO) it implies~\cite{PRD2}:
\begin{equation}
\!\!\!\!\!\left( F_V^2 -\widetilde{F}_V^2  \right) - \left( F_A^2 -\widetilde{F}_A^2 \right) \,=\, v^2\,, \label{1stWSR_LO}
\end{equation}
whereas at NLO it implies the more involved~\cite{future}
\begin{equation}
\!\!\!\!\!\left( F_V^{2} \!-\! \widetilde{F}_V^{2} \right)  \!-\! \left( F_A^{2}\!-\! \widetilde{F}_A^{2}\right)   = v^2  \left( 1\!+\! \delta_{_{\rm NLO}}^{(1)} \right) \, , \label{1stWSR_NLO}
\end{equation}
plus an additional conditions $\widetilde\delta_{_{\rm NLO}}^{(1)}=0$. Note that $\delta_{_{\rm NLO}}^{(1)}$ and $\widetilde\delta_{_{\rm NLO}}^{(1)}$ are obtained from the LO high-energy expansion of the one-loop contribution of the related correlator~\cite{ST}.
\item 2nd WSR (vanishing of the $1/s^2$ term). At LO it implies~\cite{PRD2}:
\begin{equation}
\!\!\!\!\!\left( F_V^2 -\widetilde{F}_V^2  \right) M_V^2 - \left( F_A^2 -\widetilde{F}_A^2 \right) M_A^2\,=\, 0\,. \label{2ndWSR_LO}
\end{equation}
At NLO (\ref{2ndWSR_LO}) convers into~\cite{future}
\begin{equation}
\!\!\!\!\!\!\!\!\!\!\left( F_V^{2} \!-\! \widetilde{F}_V^{2} \right) M_V^{2} - \left( F_A^{2}\!-\! \widetilde{F}_A^{2}\right) M_A^{2} = v^2 M_V^{2}  \delta_{_{\rm NLO}}^{(2)}  ,  \label{2ndWSR_NLO}
\end{equation}
plus an additional conditions $\widetilde\delta_{_{\rm NLO}}^{(2)}=0$. Note again that $\delta_{_{\rm NLO}}^{(2)}$ and $\widetilde\delta_{_{\rm NLO}}^{(2)}$ are obtained from the NLO high-energy expansion of the one-loop contribution of the related correlator~\cite{ST}.
\end{enumerate}
Whereas the 1st WSR is expected to be satisfied in gauge theories with nontrivial ultraviolet (UV) fixed points, the applicability of the 2nd WSR hinges on the nature of the UV theory under consideration. 

It is interesting to stress that in the absence of P-odd couplings (\ref{1stWSR_LO}) and (\ref{2ndWSR_LO}) together require $M_A \!>\! M_V$ and this mass hierarchy continues to hold if odd-parity couplings are assumed to be smaller than even-parity ones, $\widetilde{F}_{V,A}  \ll F_{V,A}$, which is a reasonable assumption we make.
\end{enumerate}

\section{Oblique Electroweak Observables: $S$ and $T$ at NLO}

\subsection{The observables}

We adopt here the notation used in Ref.~\cite{ST}. The calculations are carried out in the Landau gauge, which ensures that the gauge boson propagators are transverse and their self-energies,
\begin{align}
\mathcal{L}_{\mathrm{v.p.}}=& - \frac{1}{2} W^3_\mu\, \Pi^{\mu\nu}_{33}(s)W^3_\nu -\frac{1}{2}B_\mu\,\Pi^{\mu\nu}_{00}(s) B_\nu \nonumber \\ &\quad  - W^3_\mu\, \Pi^{\mu\nu}_{30}(s) B_\nu - W^+_\mu\,\Pi^{\mu\nu}_{WW}(s)W^-_\nu\,, \phantom{\frac{1}{2}}
\end{align}
can be written as
\begin{equation}
\Pi^{\mu\nu}_{ij} (q^2) \,=\, \left( -g^{\mu\nu} +
\frac{q^\mu q^\nu}{q^2}\right)\; \Pi_{ij}(q^2).
\end{equation}
The definitions of $S$ and $T$ involve $e_3$ and $e_1$, 
\begin{equation}
e_3\,=\, \Frac{g}{g'}  \; \widetilde{\Pi}_{30}(0) \, ,
\qquad
e_1\,=\,
\frac{ \Pi_{33}(0) - \Pi_{WW} (0)}{M_W^2}\,, \label{e3e1}
\end{equation}
where one has removed the tree-level Goldstone contribution from $\Pi_{30}(s)$~\cite{Peskin_Takeuchi}:
\begin{equation}
\Pi_{30}(s)\,=\,s\, \widetilde\Pi_{30}(s)\,+\,\frac{g^2 \tan{\theta_W}}{4}\, v^2 \,  .
\end{equation}
The $S$ and $T$ parameters are determined by the discrepancies between $e_3$ and $e_1$ and their respective SM contributions, denoted as $e_3^{\rm SM}$ and $e_1^{\rm SM}$, respectively:
\begin{equation}
\!\!\!\!\!S=  \Frac{16\pi}{g^2}\big(e_3 -e_3^{\rm SM}\big) ,
\quad 
T= \Frac{4\pi }{g^2 \sin^2{\theta_W}} \big(e_1-e_1^{\rm SM}\big)  .\label{STdef}
\end{equation}

\subsection{Dispersive relations}

For the calculations of $S$ and $T$ we use dispersive representations~\cite{Peskin_Takeuchi,ST}:
\begin{eqnarray}
\!\!\!\!\!S &=& \Frac{16\pi}{g^2 \tan\theta_W}\; \Int_0^\infty \Frac{{\rm ds}}{s}\; [\rho_S(s)\, -\,\rho_S(s)^{\rm SM} ] \,, \nonumber \\ 
\!\!\!\!\!T &=& \Frac{4\pi}{ g'^{\, 2} \cos^2\theta_W} \; \Int_0^\infty \Frac{{\rm ds}}{s^2} \; [\rho_T(s)\, -\, \rho_T(s)^{\rm SM}]\, , 
\label{eq:disp-rel}
\end{eqnarray}
with the spectral functions
\begin{eqnarray}
\!\!\!\!\!\rho_S(s) &=&\Frac{1}{\pi}\,\mbox{Im}\widetilde{\Pi}_{30}(s)\, , \nonumber \\
\!\!\!\!\!\rho_T(s) &=& \frac{1}{\pi}\mbox{Im}[\Sigma^{(0)}(s)-\Sigma^{(+)}(s)]\, , \label{rho}
\end{eqnarray}
being $\Sigma$ the corresponding Goldstone self-energy. At LO in $g$ and $g'$ the SM one-loop spectral function read
\begin{eqnarray}
\!\!\!\!\!\rho_S(s)^{\rm SM} \!\!&=&\!\!  \frac{g^2 \tan\theta_W}{192\pi^2} \left[ \theta(s)  \!-\!  \left(1\!-\!\frac{m_h^2}{s} \right)^3  \theta \left(s\!-\!m_h^2 \right)   \right] \! , \nonumber \\
\!\!\!\!\!\rho_T(s)^{\rm SM} \!\!&=&\!\! \Frac{3g'^{\, 2}s}{64\pi^2}\;  \bigg[-\theta(s)+\left(1-\frac{m_h^4}{s^2} \right)\theta (s-m_h^2)\bigg] \,.\nonumber\\&& \label{rho_SM}
\end{eqnarray} 

\subsection{Leading-order calculation}

At LO $T$ vanishes ($T_{\mathrm{LO}}=0$), while there is a LO contribution to $S$,
\begin{equation}
\!\!\!\!\!\!\!\left. \Pi_{30}(s) \right|_{\mathrm{LO}} =\frac{g^2  \tan{\theta_W} }{4} s  \left(\!\frac{v^2}{s}\!+\!  \frac{F_V^2\!-\!\widetilde{F}_V^2}{M_V^2\!-\!s}  \!-\! \frac{F_A^2\!-\!\widetilde{F}_A^2}{M_A^2\!-\!s} \right) . \label{PI_LO}
\end{equation}
so that, and by using (\ref{e3e1}-\ref{STdef}),
\begin{equation}
S_{\mathrm{LO}} \,=\, 4\pi  \left( \frac{F_V^2-\widetilde{F}_V^2}{M_V^2}  - \frac{F_A^2-\widetilde{F}_A^2}{M_A^2} \right)\, . \label{S_LO_0}
\end{equation}
If one assumes the 1st and the 2nd WSRs of (\ref{1stWSR_LO}) and (\ref{2ndWSR_LO}), $S_{\mathrm{LO}}$ is determined in terms of only resonance masses, 
\begin{equation}
S_{\mathrm{LO}} = 4\pi v^2 \left( \frac{1}{M_V^2} + \frac{1}{M_A^2} \right) . \label{S_LO}
\end{equation}
If only the 1st WSR is considered, and assuming $M_A > M_V$ and $\widetilde{F}_A^{2} < F_A^{2}$, (\ref{1stWSR_LO}) allows us to obtain a bound:
\begin{equation} \label{S_LO_1WSR}
\!\!\!\!\!S_{\mathrm{LO}}  = 4\pi \left\{ \!\frac{v^2}{M_V^{2}}  \! +\! \left(\!F_A^{2} \!-\! \widetilde{F}_A^{2}\! \right)\left( \!\frac{1}{M_V^{2}} \!-\! \frac{1}{M_A^{2}} \right) \right\}  >  \frac{4\pi v^2}{M_V^{2}} . \label{S_LO1WSR}
\end{equation}

\subsection{Next-to-leading-order calculation}

At NLO the number of resonance parameters increases and we have opted for an expansion in small $\widetilde{F}_{V,A}/F_{V,A}$, taking into account that odd-parity couplings are expected in our study to be smaller than even-parity couplings, that is, $\widetilde{F}_{V,A}/F_{V,A} \ll 1$. If one considers the 1st and the 2nd WSRs of (\ref{1stWSR_NLO}) and (\ref{2ndWSR_NLO}), the bosonic contributions studied here determine $S$ in terms of only $\kappa_W$, $M_{V,A}$ and the expansion parameters $\widetilde{F}_{V,A}/F_{V,A}$:
\begin{eqnarray}
\!\!\!\!\!\!\!S_{\mathrm{NLO}} \!\!\! &\!\!\!=\!\!\!&\!\!\!  4\pi v^2  \bigg(\frac{1}{M_{V}^{2}} +\frac{1}{M_{A}^{2}}\bigg) + S_{\mathrm{NLO}}^{\mathrm{P-even}} + S_{\mathrm{NLO}}^{\mathrm{P-odd}} \,, \nonumber  \\
\!\!\!\!\!\!\!S_{\mathrm{NLO}}^{\mathrm{P-even}} \!\!\!\!&\!\!\!=\!\!\!&\!\!\! \frac{1}{12\pi} \left[ \left(1-\kappa_W^2\right)  \left( \log \frac{M_V^2}{m_h^2} -\frac{11}{6} \right) \right. \nonumber \\&& \qquad   \left. +\kappa_W^2 \left( \frac{M_A^2}{M_V^2}-1 \right) \log \frac{M_A^2}{M_V^2}  \right]\,, \nonumber \\
\!\!\!\!\!\!\!S_{\mathrm{NLO}}^{\mathrm{P-odd}} \!\!\!&\!\!\!=\!\!\!&\!\!\!  \frac{1}{12\pi}  \left(\frac{\widetilde{F}_V^2}{F_V^2} +2\kappa_W^2 \frac{\widetilde{F}_V \widetilde{F}_A}{F_V F_A} - \kappa_W^2 \frac{\widetilde{F}_A^2}{F_A^2} \right) \times\nonumber \\
 &&\qquad   \left( \frac{M_A^2}{M_V^2} \!-\! 1\right)  \log \frac{M_A^2}{M_V^2} 
 \! +\! \mathcal{O}\left(\!\frac{\widetilde{F}^4_{V,A}}{F^4_{V,A}}\!\right)  \,. \label{result_SNLO}
 \end{eqnarray}
 As at LO, assuming only the 1st WSR of (\ref{1stWSR_NLO}) and $M_A \! > \! M_V$ allow us to get a lower bound for $S_{\mathrm{NLO}}$,
 \begin{eqnarray}
\!\!\!\!\!\!\!S_{\mathrm{NLO}} \!\!\! &\!\!\! > \!\!\!&\!\!\!  \frac{4\pi v^2}{M_{V}^{2}} + \Delta S_{\mathrm{NLO}}^{\mathrm{P-even}} + \Delta S_{\mathrm{NLO}}^{\mathrm{P-odd}} \,, \nonumber 
 \\
\!\!\!\!\!\!\!\Delta S_{\mathrm{NLO}}^{\mathrm{P-even}} \!\!\!\!&\!\!\!=\!\!\!&\!\!\! \frac{1}{12\pi} \left[ \left(1-\kappa_W^2\right)  \left( \log \frac{M_V^2}{m_h^2} -\frac{11}{6} \right) \right. \nonumber \\&& \qquad   \left.
 - \,\kappa_W^2\, \Bigg(\log \frac{M_A^2}{M_{V}^2}-1 + \frac{M_A^2}{M_V^2}\Bigg) \right] \,, \nonumber 
 \\
\!\!\!\!\!\!\!\Delta S_{\mathrm{NLO}}^{\mathrm{P-odd}} \!\!\!&\!\!\!=\!\!\!&\!\!\! 
\frac{1}{12\pi} \left\{  \left[ \frac{\widetilde{F}_V^2}{F_V^2}+ \kappa_W^2 \frac{\widetilde{F}_A}{F_A}\left( 2 \frac{\widetilde{F}_V}{F_V}-\frac{\widetilde{F}_A}{F_A} \right)\right] \times  \right. \nonumber 
\\
    &&\qquad   \left.\left. \left(1-\frac{M_A^2}{M_V^2}\right) \!+\! \log \frac{M_A^2}{M_V^2} \times
   \right. \right.  \nonumber 
   \\
        && \qquad \left. \left( \frac{\widetilde{F}_V^2}{F_V^2}\!-\! \kappa_W^2 \frac{\widetilde{F}_A^2}{F_A^2} \!-\! 2 \frac{\widetilde{F}_V\widetilde{F}_A}{F_VF_A}\right)
    \right\}  \!+\! \mathcal{O}\left(\!\frac{\widetilde{F}^4_{V,A}}{F^4_{V,A}}\!\right)\!.
     \nonumber \\ \label{result_SNLObis}
    \end{eqnarray}  

 Independently of the assumption about the WSRs, $T$ is determined again in terms of only $\kappa_W$, $M_{V,A}$ and $\widetilde{F}_{V,A}/F_{V,A}$:
 \begin{eqnarray}
\!\! \!\!\!\!\!T_{\mathrm{NLO}} \!\!\! &\!\!\!=\!\!\!&\!\!\!   T_{\mathrm{NLO}}^{\mathrm{P-even}} + T_{\mathrm{NLO}}^{\mathrm{P-odd}} \,, \nonumber  \\ 
\!\! \!\!\!\!\!T_{\mathrm{NLO}}^{\mathrm{P-even}}\! \!\!\! &\!\!\!=\!\!\!&\!\!\! \! \frac{3}{16\pi \!\cos^2 \!\theta_W} \!\left[ \!\left( 1\!-\!\kappa_W^2 \right) \!\left( \!1 \!-\! \log\!{\frac{M_V^2}{m_{h}^2}} \!\right)  \!+\! \kappa_W^2 \log\!{\frac{M_A^2}{M_V^2}}   \right] \!  , \nonumber \\
\!\!\!\!\!\!\!T_{\mathrm{NLO}}^{\mathrm{P-odd}} \!\!\!&\!\!\!=\!\!\!&\!\!\! \frac{3}{16\pi \!\cos^2 \!\theta_W}
\Bigg\{  2\kappa_W^2 \! \frac{\widetilde{F}_A}{F_A}\!-\! 2\frac{\widetilde{F}_V}{F_V}  \!+\!  \frac{M_V^2}{M_A^2\!-\!M_V^2} \times \nonumber \\
&&  \qquad   \log \frac{M_A^2}{M_V^2} \left( 2\frac{\widetilde{F}_V}{F_V} \!-\! 2 \kappa_W^2 \frac{M_A^2}{M_V^2} \frac{\widetilde{F}_A}{F_A}\right)    \nn \\ 
 && \qquad  
 \!+\!  \frac{M_V^2}{M_A^2\!-\!M_V^2} 
 \log \frac{M_A^2}{M_V^2} \left[\,
 \left(\kappa_W^2 \frac{\widetilde{F}_A^2}{F_A^2} \!-\! \frac{\widetilde{F}_V^2}{F_V^2}  \right) \times
\right.\nonumber \\
&& \qquad \left.
 \left(1\!+\! \frac{M_A^2}{M_V^2} \right) 
 +2 \,\frac{ \widetilde{F}_V\widetilde{F}_A}{ F_VF_A } \left( \kappa_W^2 \frac{M_A^2}{M_V^2} \!-\! 1 \right) 
  \right]\nonumber \\ &&
    \qquad+2 \left(
    \frac{\widetilde{F}_V^2}{F_V^2} \!-\!
    \!\kappa_W^2 \frac{\widetilde{F}_A^2}{F_A^2}  \!+\! \left( 1 \!-\! \kappa_W^2 \right)\!\frac{ \widetilde{F}_V\widetilde{F}_A}{ F_VF_A } \!\right)\! \Bigg\}
\nonumber \\ && \qquad  + \mathcal{O}\left(\!\frac{\widetilde{F}^3_{V,A}}{F^3_{V,A}}\!\right)
.\label{result_TNLO}  
\end{eqnarray}

\section{Phenomenology}

Now we would like to analyze the bounds on the resonance masses by comparing the theoretical determinations given in (\ref{result_SNLO}-\ref{result_TNLO}) with the experimental values of the oblique parameters. For $S$ and $T$ we consider the experimental values reported by the Particle Data Group~\cite{PDG}, $S=-0.01\pm 0.07$ and $T=0.04 \pm 0.06$ (with a correlation of $0.92$), while for the $hWW$ coupling we use the value given in Ref.~\cite{deBlas:2018tjm}, $\kappa_W=1.01\pm 0.06$. Let us remind the assumptions performed in the NLO calculation: only the lightest bosonic absorptive cuts are taken into account ($\varphi \varphi$ and $h \varphi$); odd-parity couplings are supposed to be subleading and, consequently, an expansion in $\widetilde{F}_{V,A}/F_{V,A}$ is followed; and one assumes that $M_A > M_V$ (when both WSRs are considered, this is not an assumption, but a requirement).

\subsection{Phenomenology at LO}

\begin{figure}
\begin{center}
\includegraphics[scale=0.64]{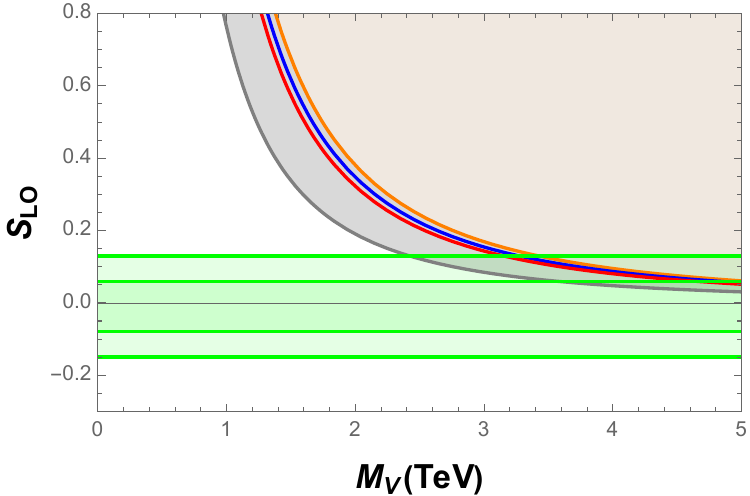}
\caption{{\small
LO predictions for $S$. The green area covers the experimentally allowed region, at 68\% and 95\% CL~\cite{PDG}. The gray region assumes the two WSRs and we indicate explicitly the corresponding lines for $M_A=M_V$ (orange), $M_A=1.1\, M_V$ (blue), $M_A=1.2\, M_V$ (red) and $M_A \to \infty$ (dark gray).  If only the 1st WSR is considered, the allowed region is given by both, the gray and the brown areas.
}}
\label{plotSLO}
\end{center}
\end{figure}

At LO $T$ vanishes and $S$ is given in (\ref{S_LO}), if one assumes both WSRs, and (\ref{S_LO1WSR}), if one considers only the 1st WSR. These results are identical to the ones we obtained in Ref.~\cite{ST}, since the inclusion of P-odd operators has not changed the LO predictions. In Figure~\ref{plotSLO} we show these predictions, together with the experimentally allowed region at 68\% and 95\% CL~\cite{PDG}. The gray area assumes both WSRs and $M_A \!>\!M_V$. The colored curves indicate explicitly the predicted results for $M_A=M_V$ (orange), $M_A=1.1\, M_V$ (blue), $M_A=1.2\, M_V$ (red) and $M_A \to \infty$ (dark gray). When only the 1st WSR is considered, the allowed range gets enlarged to the brown region. Note that the experimental data imply $M_V \!>\! 2\,$TeV.

\subsection{Phenomenology at NLO considering both WSRs}

\begin{figure}
\begin{center}
\includegraphics[scale=0.5]{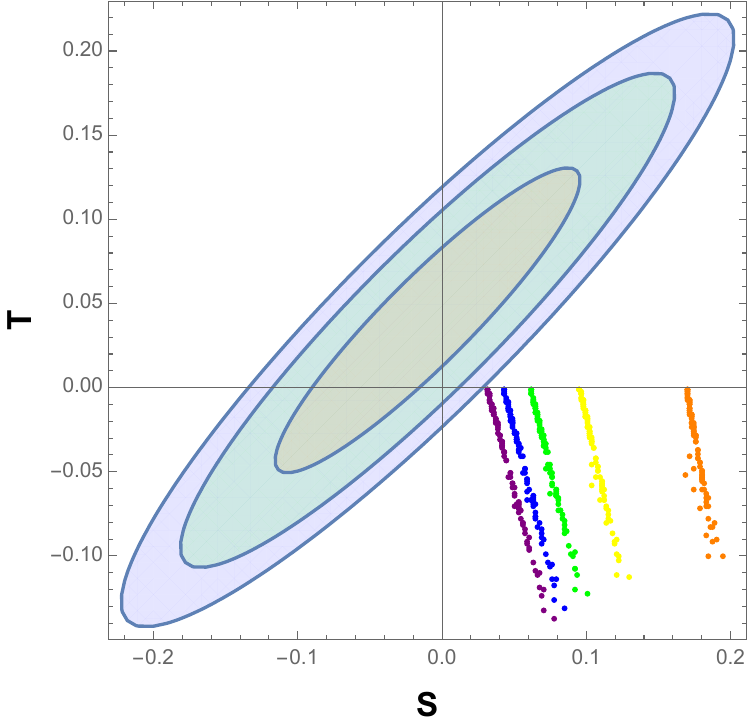} 
\caption{{\small
NLO determinations of $S$ and $T$ assuming both WSRs. The ellipses give the experimentally allowed regions of $S$ and $T$ at $68$\%, $95$\% and $99$\% CL~\cite{PDG}. The different colors of the points correspond to different values of $M_V$: $M_V=2$ (red), $3$ (orange), $4$ (yellow), $5$ (green), $6$ (blue) and $7$ (purple) TeV. The values of $\kappa_W$ and $\widetilde{F}_{V,A}/F_{V,A}$  have been generated considering normal distributions given by $\kappa_W=1.01\pm 0.06$~\cite{deBlas:2018tjm} and $\widetilde{F}_{V,A}/F_{V,A}=0.00\pm 0.33$. }}
\label{fig:NLO_2WSR}
\end{center}
\end{figure}

In this case the expressions of $S$ and $T$ are reported in (\ref{result_SNLO}) and (\ref{result_TNLO}), respectively. Note that they are given in terms of only four free parameters: $M_V$, $M_A$, $\widetilde{F}_{V}/F_{V}$ and $\widetilde{F}_{A}/F_{A}$, but the last two ones are expected to be small in our expansion and a normal distribution with $\widetilde{F}_{V,A}/F_{V,A}=0.00\pm 0.33$ is assumed. Moreover, the additional constraint $\widetilde\delta_{_{\rm NLO}}^{(2)}=0$ allows us to fix $M_A$ too (implying very close values of $M_A$ to $M_V$). In Figure~\ref{fig:NLO_2WSR} we show the results. The experimental data imply much higher values for $M_V$.

\subsection{Phenomenology at NLO considering only the 1st WSR}

\begin{figure*}
\begin{center}
\includegraphics[scale=0.42]{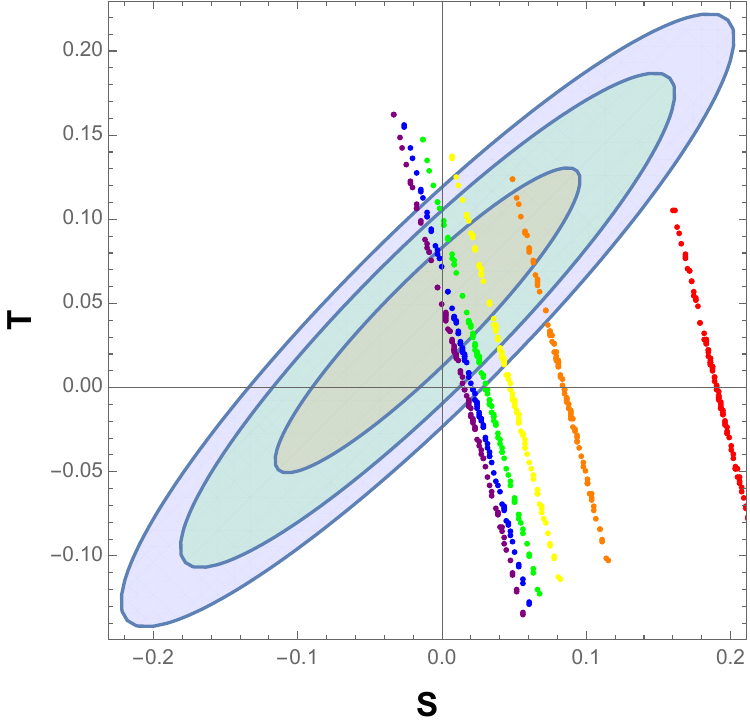} \includegraphics[scale=0.42]{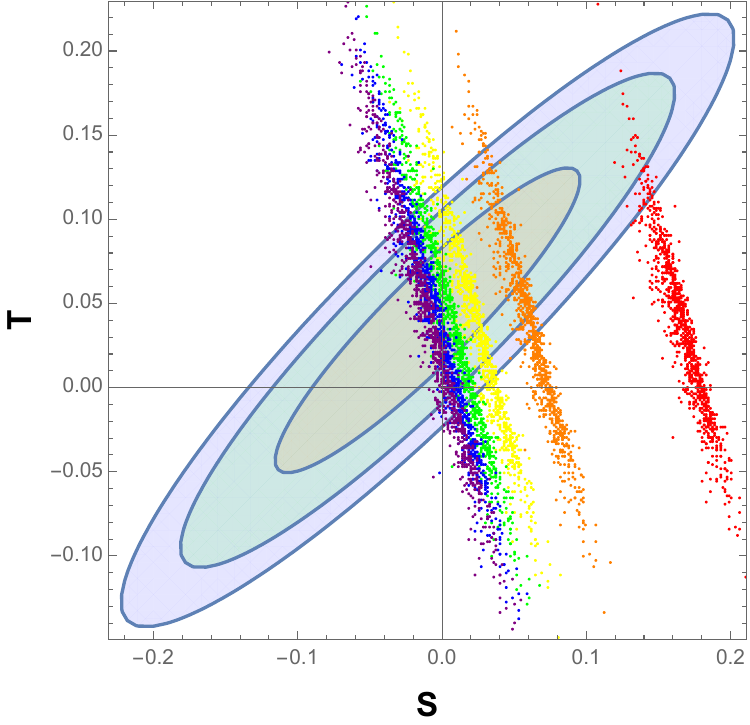} 
\includegraphics[scale=0.42]{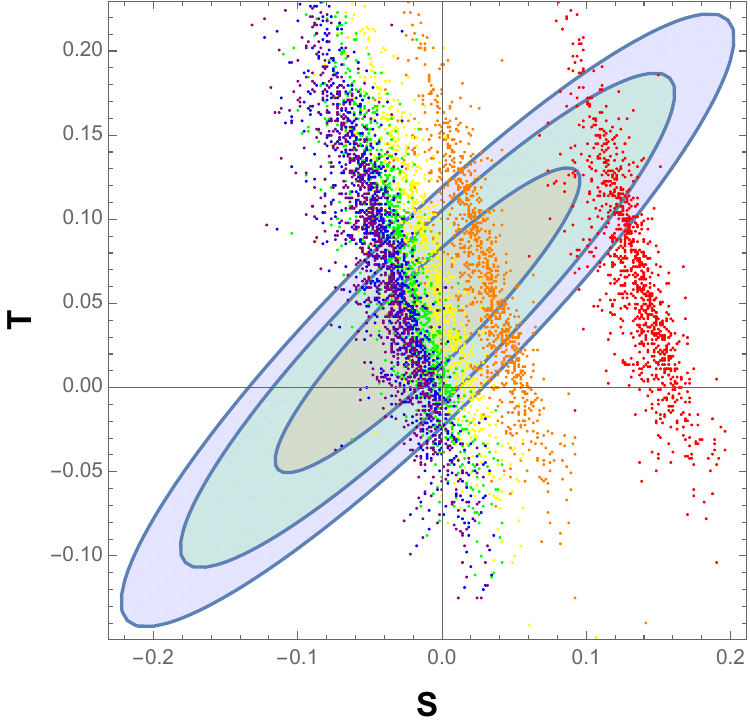}
\caption{{\small
NLO determinations of $S$ assuming only the 1st WSR, so only lower bounds of $S$ are shown. The ellipses give the experimentally allowed regions of $S$ and $T$ at $68$\%, $95$\% and $99$\% CL~\cite{PDG}. The different colors of the points correspond to different values of $M_V$: $M_V=2$ (red), $3$ (orange), $4$ (yellow), $5$ (green), $6$ (blue) and $7$ (purple) TeV. We have considered different values of $M_A$ in terms of $M_V$: $M_A=M_V$ (left), $M_A=1.2 M_V$ (centered) and $M_A=1.5M_V$ (right). The values of $\kappa_W$ and $\widetilde{F}_{V,A}/F_{V,A}$ have been generated considering normal distributions given by $\kappa_W=1.01\pm 0.06$~\cite{deBlas:2018tjm} and $\widetilde{F}_{V,A}/F_{V,A}=0.00\pm 0.33$. }}
\label{fig:NLO_1WSR}
\end{center}
\end{figure*}

Now the lower bound of $S$ in given in (\ref{result_SNLObis}) and the expresssion of $T$ is shown in (\ref{result_TNLO}). Please take note that again they are expressed using just four resonance parameters: $M_V$, $M_A$, $\widetilde{F}_{V}/F_{V}$ and $\widetilde{F}_{A}/F_{A}$. As in the previous case for $\widetilde{F}_{V}/F_{V}$ and $\widetilde{F}_{A}/F_{A}$ we consider a normal distribution with $\widetilde{F}_{V,A}/F_{V,A}=0.00\pm 0.33$.  The comparison between our determinations and the experimental values is shown in Figure~\ref{fig:NLO_1WSR}, leading to $M_V \!\gsim\! 3\,$TeV.

\section{Conclusions}

We conduct a calculation of the $S$ and $T$ oblique parameters at the next-to-leading-order level, employing an effective methodology that incorporates spin-1 heavy resonances. Our approach adopts a broad non-linear realization of Electroweak Symmetry Breaking (EWSB), devoid of any presumptions regarding the precise connection between the Higgs and the Goldstones. This results represents a first step in the update of the results reported in Ref.~\cite{ST}, encompassing a more general Lagrangian~\cite{lagrangian} and taking into account the latest experimental constraints of Ref.~\cite{PDG}. In the near future we will conclude this update by considering fermionic contributions too~\cite{future}.

It is worth highlighting that employing dispersive relations has eliminated the reliance on arbitrary cut-off values, which can introduce unphysical aspects. Additionally, we incorporate essential high-energy constraints by presuming the presence of well-behaved form factors~\cite{PRD2} and invoking the Weinberg Sum Rules (WSRs)~\cite{WSR}. This strategic approach enables us to express $S$ and $T$ using only a small set of resonance parameters: $M_V$, $M_A$, $\widetilde{F}_{V}/F_{V}$ and $\widetilde{F}_{A}/F_{A}$.

We assume that odd-parity couplings are subleading, so that we can employ an expansion in $\widetilde{F}_{V,A}/F_{V,A}$. This assumption allows us to consider a normal distribution for $\widetilde{F}_{A}/F_{A}$, with $\widetilde{F}_{V,A}/F_{V,A}=0.00\pm 0.33$. Within the context of both Weinberg Sum Rules (WSRs), $M_A$ must be a little larger than $M_V$ and our findings are graphically illustrated in Figure~\ref{fig:NLO_2WSR}. In the case where we disregard the 2nd WSR, we assume that $M_A > M_V$ and we examine the alignment between our predictions and experimental constraints in Figure~\ref{fig:NLO_1WSR}.

The primary inference drawn from our analysis is that the present electroweak precision data permits the existence of massive resonances at the natural electroweak scale, $M_R \!\gsim\! 3\,$TeV. 
Our findings align with the conclusions drawn in our earlier studies of Refs.~\cite{PRD2,ST}.

\section*{Acknowledtments}

We wish to thank the organizers for the pleasant conference. This work has been supported in part by the Spanish Government (PID2019-108655GB-I00, PID2020-114473GB-I00, PID2022-137003NB-I00); by the Generalitat Valenciana (PROMETEU/2021/071); by the Universidad Cardenal Herrera-CEU (INDI22/15); and by the ESI International Chair@CEU-UCH.

\section*{Declaration of Generative AI and AI-assisted technologies in the writing process}

During the preparation of this work the authors used ChatGPT in order to improve readability and language. After using this tool, the authors reviewed and edited the content as needed and take full responsibility for the content of the publication.


\begin{thebibliography}{99}

\bibitem{higgs}
  G.~Aad {\it et al.} [ATLAS Collaboration],
  ``Observation of a new particle in the search for the Standard Model Higgs boson with the ATLAS detector at the LHC,''
  Phys.\ Lett.\ B {\bf 716} (2012) 1
  [arXiv:1207.7214 [hep-ex]];
  S.~Chatrchyan {\it et al.} [CMS Collaboration],
  ``Observation of a New Boson at a Mass of 125 GeV with the CMS Experiment at the LHC,''
  Phys.\ Lett.\ B {\bf 716} (2012) 30
  [arXiv:1207.7235 [hep-ex]].
  
\bibitem{Buchalla:2016bse}
  G.~Buchalla, O.~Cat\`a, A.~Celis and C.~Krause,
  ``Standard Model Extended by a Heavy Singlet: Linear vs. Nonlinear EFT,''
  Nucl.\ Phys.\ B {\bf 917} (2017) 209
  [arXiv:1608.03564 [hep-ph]].

\bibitem{Pich:2018ltt}
  A.~Pich,
  ``Effective Field Theory with Nambu-Goldstone Modes,''
  Lecture Notes of the Les Houches Summer School, Vol. 108, Session CVIII (Oxford University Press, UK, 2020),
137-219 [arXiv:1804.05664 [hep-ph]].

\bibitem{lagrangian}
  A.~Pich, I.~Rosell, J.~Santos and J.~J.~Sanz-Cillero,
  ``Fingerprints of heavy scales in electroweak effective Lagrangians,''
  JHEP {\bf 1704} (2017) 012
  [arXiv:1609.06659 [hep-ph]];
  %
  C.~Krause, A.~Pich, I.~Rosell, J.~Santos and J.~J.~Sanz-Cillero,
  ``Colorful Imprints of Heavy States in the Electroweak Effective Theory,''
  JHEP {\bf 1905} (2019) 092
  [arXiv:1810.10544 [hep-ph]].

\bibitem{Peskin_Takeuchi}
  M.~E.~Peskin and T.~Takeuchi,
  ``A New constraint on a strongly interacting Higgs sector,''
  Phys.\ Rev.\ Lett.\  {\bf 65} (1990) 964;
  ``Estimation of oblique electroweak corrections,''
  Phys.\ Rev.\ D {\bf 46} (1992) 381.

\bibitem{ST}
  A.~Pich, I.~Rosell and J.~J.~Sanz-Cillero,
  ``Viability of strongly-coupled scenarios with a light Higgs-like boson,''
  Phys.\ Rev.\ Lett.\  {\bf 110} (2013) 181801
  [arXiv:1212.6769 [hep-ph]];
%
  ``Oblique S and T Constraints on Electroweak Strongly-Coupled Models with a Light Higgs,''
  JHEP {\bf 1401} (2014) 157
  [arXiv:1310.3121 [hep-ph]].
 
   \bibitem{future}
A.~Pich, I.~Rosell and J.~J.~Sanz-Cillero, work in preparation.

   \bibitem{Weinberg}
  S.~Weinberg,
  ``Phenomenological Lagrangians,''
  Physica A {\bf 96} (1979) no.1-2,  327.
  
\bibitem{PRD2}
A.~Pich, I.~Rosell, J.~Santos and J.~J.~Sanz-Cillero,
  ``Low-energy signals of strongly-coupled electroweak symmetry-breaking scenarios,''
  Phys.\ Rev.\ D {\bf 93} (2016) no.5,  055041
  [arXiv:1510.03114 [hep-ph]];
A.~Pich, I.~Rosell and J.~J.~Sanz-Cillero,
``Bottom-up approach within the electroweak effective theory: Constraining heavy resonances,''
Phys. Rev. D \textbf{102} (2020) no.3, 035012
[arXiv:2004.02827 [hep-ph]].

\bibitem{Bernard:1975cd}
  C.~W.~Bernard, A.~Duncan, J.~LoSecco and S.~Weinberg,
  ``Exact Spectral Function Sum Rules,''
  Phys.\ Rev.\ D {\bf 12} (1975) 792.

      \bibitem{WSR}
  S.~Weinberg,
  ``Precise relations between the spectra of vector and axial vector mesons,''
  Phys.\ Rev.\ Lett.\  {\bf 18} (1967) 507.

\bibitem{PDG}
R.~L.~Workman \textit{et al.} [Particle Data Group],
``Review of Particle Physics,''
PTEP \textbf{2022} (2022), 083C01

\bibitem{deBlas:2018tjm}
  J.~de Blas, O.~Eberhardt and C.~Krause,
  ``Current and Future Constraints on Higgs Couplings in the Nonlinear Effective Theory,''
  JHEP {\bf 1807} (2018) 048
  [arXiv:1803.00939 [hep-ph]].


  \end{thebibliography}
\end{document}